\documentclass[preprintnumbers,aps,prl,groupedaddress,twocolumn]{revtex4}
\usepackage{multirow}
\usepackage{hyperref}
\usepackage{mathrsfs}
\usepackage{amsmath}
\usepackage{graphicx}
\usepackage{color}

\hypersetup{
	colorlinks=true,
	linkcolor=blue,
	filecolor=blue,      
	urlcolor=blue,
	citecolor=blue,
}

\begin{document}
\draft
\title{Characterization of non-Markovianity with maximal extractable
qubit-reservoir entanglement}
\author{Pei-Rong Han$^{1,2}$}
\thanks{These authors contribute equally to this work.}
\author{Fan Wu$^{2}$}
\thanks{These authors contribute equally to this work.}
\author{Xin-Jie Huang$^{2}$}
\thanks{These authors contribute equally to this work.}
\author{Huai-Zhi Wu$^{2}$, Wei Yi$^{3,4}$}
\author{Jianming Wen$^{5,6,\dagger}$}
\author{Zhen-Biao Yang$^{2,7,\dagger}$}
\author{Shi-Biao Zheng$^{2,7,}$}
\thanks{Authors to whom correspondence should be addressed: jianming.wen@gmail.com, zbyang@fzu.edu.cn, t96034@fzu.edu.cn}
\address{$^{1}$School of Physics and Mechanical and Electrical Engineering, Longyan University, Longyan 364012, China\\
$^{2}$Fujian Key Laboratory of Quantum Information and Quantum Optics, College of Physics and Information Engineering, Fuzhou University,\\
Fuzhou 350108, China\\
$^{3}$CAS Key Laboratory of Quantum Information, University of Science and\\
Technology of China, Hefei 230026, China\\
$^{4}$CAS Center for Excellence in Quantum Information and Quantum Physics,\\
University of Science and Technology of China, Hefei 230026, China\\
$^{5}$Department of Electrical and Computer Engineering, Binghamton University, Binghamton, New York 13902, USA\\
$^{6}$Department of Physics, Kennesaw State University, Marietta, Georgia 30060, USA\\
$^{7}$Hefei National Laboratory, Hefei 230088, China}
\date{\today }
\begin{abstract}
Understanding the dynamical behavior of a qubit in a reservoir is critical
to applications in quantum technological protocols, ranging from quantum
computation to quantum metrology. The effect of the reservoir depends on
reservoir's spectral structure, as well as on the qubit-reservoir coupling
strength. We here propose a measure for quantifying the non-Markovian effect
of a reservoir with a Lorentzian spectrum, based on the maximum
qubit-reservoir quantum entanglement that can be extracted. Numerical
simulation shows this entanglement exhibits a monotonous behavior in
response to the variation of the coupling strength. We confirm the validity
of this measure with an experiment, where a superconducting qubit is
controllably coupled to a lossy resonator, which acts as a reservoir for
the qubit. The experimental results illustrate the maximal extractable
entanglement is progressively increased with the strengthening of the
non-Markovianity.
\end{abstract}

\vskip0.5cm

\maketitle

One of the most fundamental postulations of quantum mechanics is that the
evolution of an isolated quantum system is governed by the Schr\"odinger
equation. However, any system is inevitably coupled to its environment,
which can be regarded as a reservoir containing many bosonic field modes
\cite{1,2,3,4,5}. According to the spectral structure, reservoirs can be classified
into Markovian and non-Markovian. The Markovian reservoir has a flat
spectrum, which makes it lose the memory about the past history.
Consequently, the leakage of information and energy from the system to the
environment becomes irreversible. Due to this irreversibility, the system's
dynamics, obtained by tracing over the reservoir modes, is governed by the
Lindblad master equation \cite{1}. The resulting time evolution displays a
monotonous decaying feature, without any flow of information from the
reservoir back to the system.

When the spectral structure of the reservoir is modified, it can become
non-Markovian. 
Due to this spectral modification, the reservoir exhibits a memory effect, which is characterized by the future state of the reservoir depending on its history. This memory effect implies that the information flow from the system to the reservoir becomes partly irreversible. The degree of non-Markovianity is quantified by the extent to which information can flow back from the reservoir to the system. This reversibility is influenced by the ratio between the system-reservoir coupling strength and the spectral width of the reservoir. A higher ratio indicates a greater ability of the reservoir to return information to the system, thereby enhancing the non-Markovian behavior.
Decaying resonators represent the most typical type of
non-Markovian reservoirs, which have a Lorentzian spectrum \cite{1,2}. The memory
effect of the reservoir depends on the ratio of the system-reservoir
coupling to the spectral linewidth. When this ratio tends to infinity, the
spectral spread can be neglected, so that the reservoir reduces to a
single-mode resonator, which exchanges energy with the system in a
reversible manner. In the opposite limit, the spectral spread can be considered
infinite, so that the reservoir is approximately Markovian. In the intermediate
regime, the reservoir displays some memory effect, which is gradually lost
with the decrease of the coupling strength.

In the past decade, many efforts have been devoted to the quantification of
non-Markovianity \cite{6,7,8,9,10, 11,12,13,14}. It has been shown that the quantum memory effect
can be quantified by tracking the evolutions of two initially distinct
quantum states in the reservoir \cite{6}. When the reservoir is Markovian, the
distinguishability between these two states is monotonously reduced and
finally tends to zero when the information is completely leaked into the
environment. For a non-Markovian process, this distinguishability displays
an oscillatory pattern, as a consequence of the reversed information flow
from the reservoir back to the system. Alternatively, the non-Markovianity
can also be measured by using an ancilla that is initially entangled with
the system in the reservoir, 
and tracking the ancilla-system entanglement during the system-reservoir interaction \cite{7,11,12,13,14}. The non-Markovian degree is quantified by the extent of temporal increase in the entanglement between the open system and the ancilla.
The reservoir's memory effect is characterized by the extent to which the lost entanglement can be revived. Based on these measures, the non-Markovian effects have been
experimentally explored in different systems, including photonic qubits
\cite{15,16,17}, nitrogen-vacancy centers \cite{18,19,20}, and ion traps \cite{21}. Recently,
Gaikwad et al. explored the crossover between the Markovian and non-Markovian
dynamics with a superconducting processor \cite{22}, where the entanglement
between a test qubit and an ancilla is used as a measure to quantify the
non-Markovianity of the environment of the test qubit, comprised of a third
qubit and its readout resonator.

We here propose an alternative measure for quantifying the memory effect for
a general non-Markovian reservoir that interacts with a qubit, and present
an experimental demonstration. Instead of introducing an ancilla 
that is initially entangled with the test qubit, the non-Markovianity is directly quantified by the
extractable entanglement between the qubit and the reservoir generated by
their interaction. When the qubit is initially in the excited state, the
excitation is shared by the system and the reservoir due to their
interaction. This excitation sharing results in the qubit-reservoir
entanglement. The amount of the extractable maximum entanglement depends the
non-Markovianity of the reservoir. In the strong coupling regime, the
system-reservoir dynamics approximately reduces to the vacuum Rabi
oscillation of the Jaynes-Cummings (JC) model, where the qubit-photon
entanglement is completely extractable, with the maximum value approximating
1. With the decrease of the coupling strength, the amount of the extractable
maximum entanglement is gradually reduced due to the spreading of the
spectrum. In the Markovian limit, no entanglement can be extracted. The
experimental data used to demonstrate the crossover between the Markovian
and non-Markovian dynamics is taken from a previous experiment \cite{23}, where a
non-Hermitian JC model was synthesized with a superconducting qubit
controllably coupled to its readout resonator. Here the readout resonator is
taken as a reservoir. The experimental results confirm the theoretical
predictions.
Compared to previous measures, the present measure offers significant advantages. Notably, it does not require precise comparison between the evolution trajectories of two initially distinct quantum states, as referenced in \cite{6}. Additionally, our method eliminates the need for any ancilla qubit, which has been a requisite in other approaches \cite{7,22}.

The system under consideration is composed of a qubit of frequency $\omega
_{0}$ coupled to a non-Markovian reservoir whose spectrum is a continuum. The
dynamics of the entire system, in the framework rotating at the qubit
frequency, is governed by the Hamiltonian

\begin{equation}
H=\left\vert g\right\rangle \left\langle e\right\vert 
\displaystyle\int %
\limits_{0}^{\infty }d\omega J(\omega )\lambda (\omega )e^{i\delta (\omega
)t}a^{\dagger }(\omega )+H.c., 
\end{equation}
where $\left\vert e\right\rangle $\ and $\left\vert g\right\rangle $\ respectively denote
the excited and ground state of the qubit, $J(\omega )$ is the spectral
density of the reservoir, and $a^{\dagger }(\omega )$\ and $a(\omega )$\ are respectively
the creation and annihilation operators for the bosonic mode with the
frequency $\omega $ in the reservoir, which is coupled to the qubit with a
coupling strength $\lambda (\omega )$\ and a detuning $\delta (\omega )$.
For a leaky resonator, the spectral shape is Lorentzian, 
\begin{equation}
J(\omega )=\frac{1}{\pi }\frac{\kappa /2}{(\omega -\omega _{0})^{2}+(\kappa
/2)^{2}}, 
\end{equation}
where $\kappa $ represents the spectral line-width, which depends on the
quality and the central frequency $\omega _{0}$ of the resonator. In
conventional cavity quantum electrodynamics (QED) and circuit QED, cavities
or resonators used to hold photonic modes can be modeled as such reservoirs
\cite{1}.

Such a non-Markovian reservoir is equivalent to the combination of a single
lossy photonic mode of frequency $\omega _{0}$ and a Markovian
reservoir, into which the photonic mode continuously leaks its energy 
with a rate $\kappa $. Hereafter, we will use the symbols ``$Q$", ``$M_{0}$",
and ``$R_{MR}$" to denote, respectively, the qubit, the photonic mode, and the
Markovian reservoir, which corresponds to a flat continuum. When $Q$
is initially in the excited state $\left\vert e\right\rangle $ and $M_{0}$
in vacuum state $\left\vert 0\right\rangle $, after an interaction time $t$,
the qubit, the photonic mode, and the Markovian reservoir are evolved to the state
\begin{equation}
\left\vert \Phi (t)\right\rangle =\left\vert \psi (t)\right\rangle
\left\vert 0\right\rangle_{MR}+\sqrt{1-\left\langle \psi (t)| \psi (t)\right\rangle }\left\vert g,0\right\rangle \left\vert
1\right\rangle_{MR}, 
\end{equation}
where $\left\vert 0\right\rangle_{MR}$ and $\left\vert 1\right\rangle_{MR}$
denote the zero- and one-photon state for the Markovian reservoir,
respectively. When $\xi =4\lambda _{0}/\kappa >1$, $\left\vert \psi
(t)\right\rangle $ is given by 
\begin{eqnarray}
\begin{split}
\left\vert \psi (t)\right\rangle =  \;&e^{-\kappa t/4}\{[\cos (\Omega t)+\frac{%
\kappa }{4\Omega }\sin (\Omega t)]\left\vert e,0\right\rangle \\
&-i\frac{\lambda _{0}}{\Omega }\sin (\Omega t)]\left\vert g,1\right\rangle
\},
\end{split}
\end{eqnarray}%
where $\Omega =\frac{\kappa }{4}\sqrt{\xi ^{2}-1}$ with $\lambda
_{0}=\lambda (\omega _{0})$. For $\xi <1$,
the trigonometric functions are replaced by the corresponding hyperbolic
Functions, with $\Omega =$ $\sqrt{\kappa ^{2}/16-\lambda _{0}^{2}}$.

When the photon is leaked into the Markovian reservoir, it cannot be
re-absorbed by the qubit and the information is completely lost. Once this
event occurs, both $Q$ and $M_{0}$ jump to their ground state. The
correlation between the Markovian reservoir and the qubit or the bosonic
mode cannot be extracted. The part of quantum correlation that can be
extracted is that associated with the non-unitarily evolved wavefunction $
\left\vert \psi (t)\right\rangle $. This entanglement is characterized by the concurrence
\cite{24}
\begin{equation}
{\cal C}=\left\langle \psi (t)| \psi (t)\right\rangle \sin
(2\theta ), 
\label{C}
\end{equation}%
where%
\begin{equation}
\theta =\arctan \left\vert \frac{4\lambda _{0}\sin (\Omega t)}{4\Omega \cos
(\Omega t)+\kappa \sin (\Omega t)}\right\vert . 
\end{equation}%
The concurrence ${\cal C}$ measures the unconditional entanglement between $Q$ and $M_{0}$,
averaged over the trajectories with and without quantum jumps.
We note that this concurrence quantifies the amount of the qubit-reservoir entanglement that can be extracted. For a Markovian reservoir with $\kappa\rightarrow\infty$, the entanglement between the system and reservoir cannot be extracted, resulting in an extractable system-reservoir entanglement of zero. This implies that such a concurrence acts as a witness of non-Markovianity \cite{11}.

\begin{figure}[htbp]
	\centering
	\includegraphics[width=3.4in]{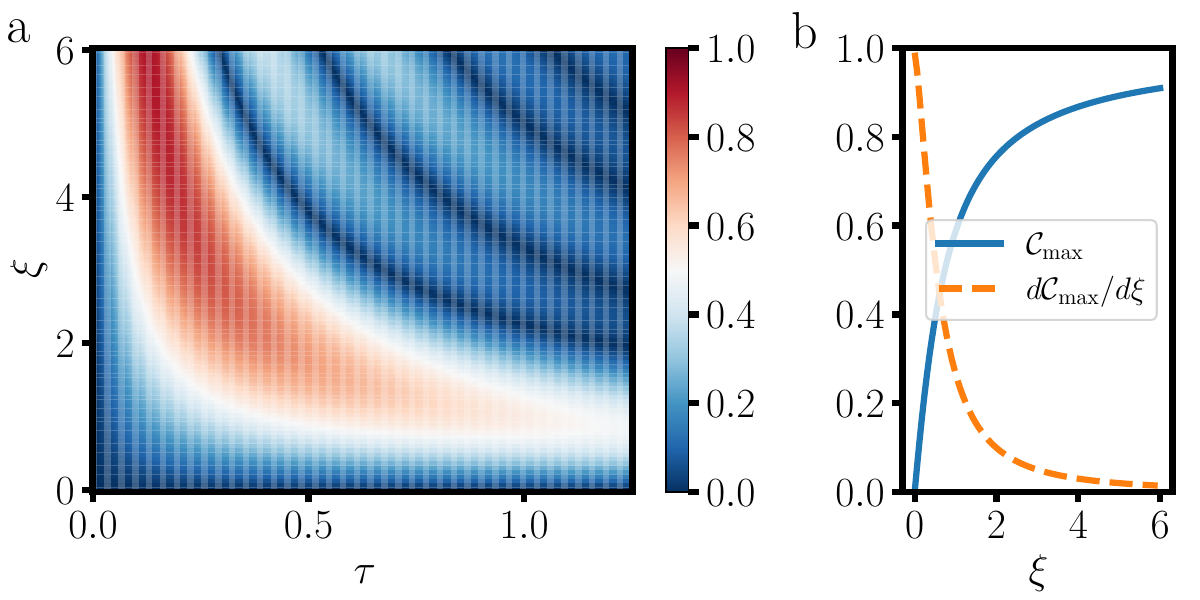}
	\caption{Theoretical results. \textbf{a} Extractable concurrence ($%
		{\cal C}$) between a qubit and a reservoir of photonic modes versus $\xi $
		and $\tau $. The reservoir, initially in the vaccum state, is composed of
		photonic modes with a Lorentzian spectrum with a line width $\kappa $. The
		qubit, initially in the excited state, is on resonance with the central
		frequency of the continuum, $\omega _{0}$. Here $\tau =\kappa t/4$ denotes
		the rescaled interaction time, and $\xi $ is the ratio of the effective
		coupling strength between the qubit and the photonic mode with frequency $\omega _{0}$ to $\kappa /4$, and $\tau $. \textbf{b} Maximum of $\mathcal{C}$ during the evolution versus $\xi$. The dashed line is the derivative of $\mathcal{C}_{\max}$ with respect to $\xi$.}

	\label{Fig1}
\end{figure}

To illustrate the relation between the extractable quantum entanglement and
the non-Markovianity of the system dynamics, we display ${\cal C}$ versus $%
\xi$ and the rescaled interaction time $\tau =\kappa t/4$ in Fig. \ref{Fig1}a. When $\xi \rightarrow \infty $, the line width of the Lorentzian
spectrum of the reservoir tends to 0 as compared to the qubit-reservoir
coupling strength, so that the reservoir can be approximately regarded as a
single photonic mode. Consequently, ${\cal C}$ nearly exhibits an
oscillatory pattern with the period equal to half of that of the vacuum Rabi
oscillation arising from the $Q$-$M_{0}$ swapping coupling, $\pi /\lambda
_{0}$. ${\cal C}$ approaches the maximum value $1$ at the time $t=\pi
/(4\lambda _{0})$, where the photon is equally shared by $Q$ and $M_{0}$.
With the decrease of $\xi $, the line width of the reservoir cannot
neglected. In this case, the photon swapping is only partly irreversible.
Due to the progressive leakage of the information into $R_{MR}$, the local
maximum of ${\cal C}$ gradually decreases. For $\xi \ll 1$, ${\cal C}$
almost remains to be $0$, which can be interpreted as follows. In this weak
coupling regime, the photon is immediately leaked to the Markovian reservoir
as long as it is transferred from the qubit to the $M_{0}$ due to the strong
dissipation. As a consequence, $M_{0}$ remains unpopulated, but mediates a
coupling between $Q$ and $R_{MR}$, which results a completely irreversible
dissipation process with a rate of $4\lambda _{0}^{2}/\kappa $.

To reveal how ${\cal C}$ can quantitatively manifest the extent of the
non-Markovianity of the reservoir, we calculate the maximal value of ${\cal C%
}$ (${\cal C}_{\max }$) during the evolution for different values of $\xi $.
For a given $\xi $, the time for the maximization of ${\cal C}$ is obtained by
setting $d{\cal C}/dt =0$. The optimal time depends upon $\Omega(\xi)$ as
\begin{equation}
t_{opt} = \left|\frac{2\arctan{\left(\frac{1}{2}\sqrt{\frac{\kappa^2+12\Omega^2-4\lambda_0\sqrt{\kappa^2+8\Omega^2}}{\Omega^2}}\right)}}{\Omega}\right|.
\end{equation}
Substituting $t_{opt}$ into Eq. (\ref{C}), we obtain the maximally extractable concurrence,
${\cal C}_{\max}$. We show ${\cal C}_{\max}$ as a function of $\xi$ in Fig. \ref{Fig1}b (solid line), which clearly shows that ${\cal C}_{\max}$ monotonously increases with $\xi$. The dashed line represents $d{\cal C}_{\max}/d\xi$ versus $\xi$, which illustrates that the changing rate of ${\cal C}_{\max}$ decreases with the increase of $\xi$, tending to 0 in the limit $\xi\rightarrow \infty$, where ${\cal C}_{\max}$ is saturated to 1. This implies that ${\cal C}_{\max }$
can serve as a measure for quantifying the non-Markovianity of the reservoir of bosonic modes with a continuum.

\begin{figure}[htbp]
	\centering
	\includegraphics[width=3in]{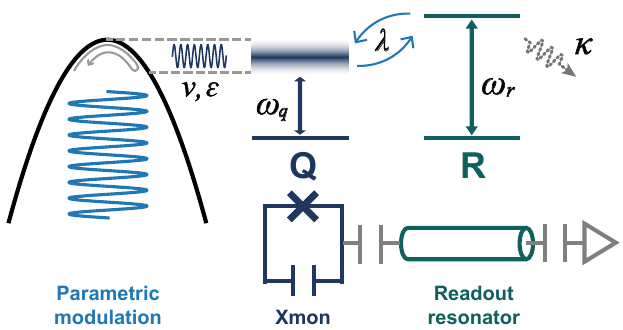}
	\caption{Schematic of the experimental system. The system consists of a Josephson-based artificial atom, known as the Xmon qubit ($Q$), and its associated readout resonator ($R$), which has a decay rate of $\kappa=5$ MHz. This resonator functions as a non-Markovian reservoir for $Q$. The coupling between $Q$ and $R$ is facilitated by a sideband modulation applied parametrically to $Q$.
	}
	\label{Fig2}
\end{figure}

To confirm the validity of the approach, we use the data measured in a
previous experiment focusing on the exceptional entanglement phase
transition in a non-Hermitian qubit-photon model \cite{23}. The experiment was
performed with a circuit electrodynamics system (Fig.~\ref{Fig2}), where an Xmon qubit with a
negligible dissipative rate was coupled to its readout resonator with a
non-negligible photonic decaying rate $\kappa =5$ MHz through a sideband
interaction, mediated by a parametric modulation with the amplitude $\varepsilon$ and frequency $\nu$. Here the readout resonator
is taken as the reservoir for the qubit, other than a part of the system.
When $\nu=(\omega_r-\omega_q)/n$, the qubit is effectively coupled to the resonator at the $n$th upper sideband, with an effective coupling strength $\lambda=gJ_n(\varepsilon/\nu)$. Here, $\omega_q$ and $\omega_r$ denote the frequencies of the qubit and the resonator, respectively, $g$ stands for their on-resonance coupling strength, and $J_n(\mu)$ is the $n$th-order Bessel function of the first kind \cite{25,26,27}. In our experiment, the effective qubit-photon coupling strength is carefully controlled by tuning the modulating frequency and amplitude. To minimize the crosstalk, we have employed first- and second-order sideband couplings in the designated regions $0<\xi<1$ and $\xi>1$, respectively. This strategic approach ensures that the system operates with the desired precision and reduces unintended interactions.
In the previous work we display the evolution of the
qubit-photon entanglement associated with the no-jump trajectory, calculated with the output qubit-photon state projected onto the single-excitation subspace $\left\{ \left\vert e,0\right\rangle ,\left\vert g,1\right\rangle \right\} $. Such a conditional output state was obtained by reconstructing the qubit-photon density matrix with the assistance of an ancilla qubit, and then discarding the element associated to $\left\vert g,0\right\rangle$.
The evolution of the post-selected output state coincides with the no-jump
trajectory governed by the non-Hermitian Hamiltonian. On the contrary,
as discussed above, the non-Markovianity is manifested here
by the entanglement evolution associated with the whole output density matrix, 
reconstructed in the subspace $\left\{
\left\vert e,0\right\rangle ,\left\vert g,1\right\rangle ,\left\vert
g,0\right\rangle \right\} $.

\begin{figure}[htbp]
	\centering
	\includegraphics[width=3.4in]{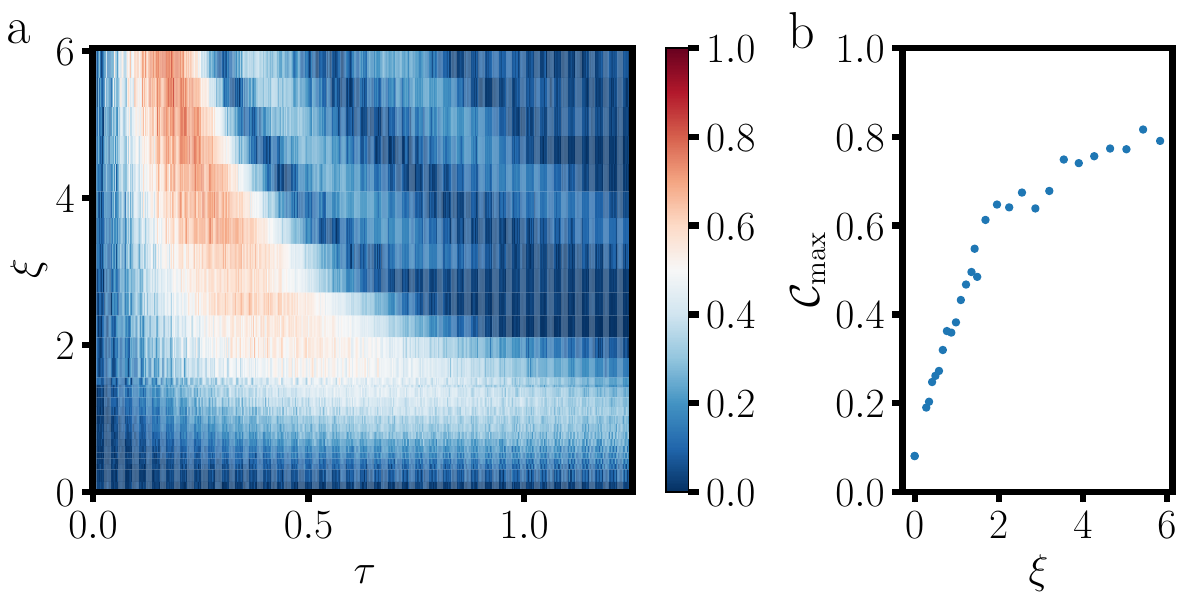}
	\caption{Experimental results. \textbf{a} Measured ${\cal C}$ versus $%
		\xi $ and $\tau $. The data is taken from a previous experiment \cite{23}, which
		was performed in a circuit QED system, where the readout resonator of an Xmon
		qubit with a photonic decaying rate $\kappa $ is taken as the reservoir. \textbf{b}
		Measured ${\cal C}_{\max }$ versus $\xi $. 
	}
	\label{Fig3}
\end{figure}

Fig. \ref{Fig3}a presents the measured concurrence, without post-selection, as a
function of $\xi $ and $t$. As expected, in the strong coupling limit $\xi
\gg 1$, the unconditional qubit-photon concurrence displays nearly perfect
periodic oscillations, which can be explained as follow. For a single
photonic mode with a detuning $\delta (\omega )=\omega -\omega _{0}$, the
oscillating frequency associated with the excitation swapping is $\chi
_{\omega }=2\sqrt{\lambda ^{2}(\omega )+\delta ^{2}(\omega )/4}$. We here
set $\left\vert \delta (\omega )\right\vert \ll \omega $ so that $\lambda
(\omega )\simeq \lambda _{0}$. When the line-width of the Lorentzian
spectrum of the resonator is much smaller than the coupling strength $%
\lambda _{0}$, i.e., $\left\vert \delta (\omega )\right\vert \ll \lambda
_{0} $, we have $\chi _{\omega }\simeq 2\lambda _{0}$. This implies that all
the bosonic modes within the spectral width of the reservoir are almost
equally coupled to the qubit, keeping synchronous during interaction with
the qubit, thereby collectively behaving like a single mode. With the
decrease of $\xi $, the oscillatory feature is progressively weakened owing
to the non-negligible spectral width.

Fig. \ref{Fig3}b displays the maximum of the measured qubit-photon concurrence during
the interaction, as a function of $\xi $. The slight fluctuations are mainly
due to the imperfect control of the effective qubit-photon coupling strength.
The results confirm
that ${\cal C}_{\max }$ progressively increases with $\xi $, tending to 1 in
the strong coupling regime. The monotonousness of ${\cal C}_{\max }$ implies
that it can serve as a measure for characterizing the non-Markovianity.

Due to unavoidable experimental imperfections, the measured concurrences are slightly lower than the ideal values depicted in Fig. \ref{Fig1}. The primary source of error arises from the influence of neglected off-resonant terms in the expansion of the parametric modulation. These off-resonant terms cause the dynamics of the experimental system to diverge from the idealized model. Specifically, the dominant off-resonant term is the carrier coupling with a strength of $gJ_{0}(\varepsilon/\nu)$, which introduces fast, slight oscillations in the system’s behavior. This effect is consistent with the observation detailed in \cite{23}.
However, we note that parametric modulation becomes unnecessary when the resonator frequency falls within the adjustable range of the qubit frequency. In such cases, the unwanted off-resonant terms mentioned earlier are effectively avoided. Additionally, the impact of qubit frequency fluctuations is mitigated by the qubit-resonator interaction, which inherently enhances the system’s robustness against dephasing noise, as explained below. Specifically, the qubit-resonator interaction generates two dressed states separated by an energy gap, thereby preventing dephasing noise from inducing transitions between these dressed states \cite{28}.

In conclusion, we have shown that the maximal entanglement between a qubit
and its reservoir extractable can be used as a measure for
quantifying the non-Markovianity of the dynamics. 
When the qubit-reservoir coupling is much smaller than the spectral line-width of the reservoir, the leakage of the excitation from the qubit to the reservoir is almost irreversible. Consequently, most of the qubit-reservoir entanglement produced by their interaction is unextractable, reflecting the Markovian feature.
With the increase of the coupling, the reservoir displays
a non-Markovian effect, manifested by the nonvanishing qubit-reservoir
entanglement that can be extracted. We illustrate this measure with a
circuit QED experiment, where the readout resonator of a qubit acts as a
reservoir of photonic modes with a continuous spectrum. By tuning the
effective qubit-resonator coupling, we observed the progressive crossover
from the Markovian to non-Markovian dynamics.

This work was supported by the National Natural Science Foundation of China
(Grant Nos. 12274080, 12474356, 12475015, 11875108, 12174058, 12204105, and 12374479), the NSF (Grant No. 2329027) and Innovation Program for Quantum Science and
Technology (Grant No. 2021ZD0300200).


\end{document}